\documentclass{ws-ijmpd}                                                        
  
\input cyracc.def

\def\Sigmakin{{}^{\rm k}\Sigma_\alpha}
\def\kappabf{\mbox{\boldmath$\kappa$\unboldmath}}

\begin{document}


\markboth{G.F.\ Rubilar, Yu.N.\ Obukhov, F.W.\ Hehl}
{Generally covariant Fresnel equation and the emergence of the
  light cone structure}
 
%
\catchline{}{}{}
%
 
\title{GENERALLY COVARIANT FRESNEL EQUATION\\ AND THE EMERGENCE OF THE 
  LIGHT CONE STRUCTURE\\ IN LINEAR PRE-METRIC ELECTRODYNAMICS}

\author{Guillermo F.\  Rubilar\footnote{Email: gr@thp.uni-koeln.de.},
  Yuri N.\ Obukhov\footnote{Email: yo@thp.uni-koeln.de. Permanent
    address: Department of Theoretical Physics, Moscow State
    University, 117234 Moscow, Russia.} , and 
Friedrich W.\ Hehl\footnote{Email: hehl@thp.uni-koeln.de.}}

\address{Institute for Theoretical Physics, University of Cologne\\ 
50923 K\"oln,  Germany}

\maketitle
\pub{Received (received date)}{Revised (revised date)}                         

\begin{abstract}  
  We study the {\em propagation of electromagnetic waves} in a 
  spacetime devoid of a metric but equipped with a {\em linear} 
  electromagnetic spacetime relation $H\sim\chi\cdot F$. Here $H$ is 
  the electromagnetic excitation $({\cal D},{\cal H})$ and $F$ the 
  field strength $(E,B)$, whereas $\chi$ (36 independent components) 
  characterizes the electromagnetic permittivity/permeability of 
  spacetime. We derive analytically the corresponding Fresnel equation 
  and show that it is always quartic in the wave covectors. We study 
  the `Fresnel tensor density' ${\cal G}^{ijkl}$ as (cubic) function 
  of $\chi$ and identify the leading part of $\chi$ (20 components) as 
  indispensable for light propagation. Upon requiring 
  electric/magnetic reciprocity of the spacetime relation, the leading 
  part of $\chi$ induces the {\em light cone} structure of spacetime 
  (9 components), i.e., the spacetime metric up to a function. The 
  possible existence of an Abelian {\em axion} field (1 component of 
  $\chi$) and/or of a {\em skewon} field (15 components) and their 
  effect on light propagation is discussed in some detail. The newly 
  introduced skewon field is expected to be T-odd and related to 
  dissipation.  {\em file nonsym37.tex, 2002-03-24} 
\end{abstract}  
 
 
\section{Introduction}  
  
In pre-metric electrodynamics, 
the axioms of electric charge and of magnetic flux conservation 
manifest themselves in the Maxwell equations for the excitation 
$H=({\cal D},{\cal H})$ and the field strength $F=(E,B)$, see 
\cite{gentle}: 
\begin{equation}\label{me} 
dH=J, \qquad dF=0 \,. 
\end{equation} 
Here $J=(\rho,j)$ is the electric current.  These equations keep their 
form in all reference frames and, since they are metric independent, 
in all gravitational fields as well. In order to develop this scheme 
into a predictive physical theory, one needs additionally a {\it 
  spacetime relation} $H= H(F)$, which is a functional in general but 
is assumed here to be of a local nature.  It should be well 
understood, this spacetime relation is a {\it universal} constitutive law 
for the {\it vacuum}. We do {\it not} treat the constitutive behavior 
of material media here, we rather search for a universal law for the 
vacuum, i.e., for the spacetime manifold 
itself\,\footnote{Nevertheless, since Maxwell's equations for material 
  media have the same appearance as the vacuum equations (\ref{me}), a 
  constitutive relation for a material medium can be reminiscent of 
  the {\it spacetime relation} --- and we will profit from this 
  analogy.}. 
 
The simplest assumption is that the spacetime relation is {\em local 
  and linear}. If local coordinates $x^i$ are given, with 
$i,j,... =0,1,2,3$, we can decompose the excitation and field strength 
2-forms into their components according to 
\begin{equation}   
H = {\frac 1 2}\,H_{ij}\,dx^i\wedge dx^j,\qquad   
F = {\frac 1 2}\,F_{ij}\,dx^i\wedge dx^j.\label{geo1}   
\end{equation}  
Then the {\em linear} spacetime relation, see \cite{Post,HO02}, can be 
written as 
\begin{equation}\label{cl}   
H_{ij}=\frac{1}{4}\,{\hat \epsilon}_{ijkl}\,\chi^{klmn}\,F_{mn} \,.  
\end{equation}   
Here ${\hat\epsilon}_{ijkl}$ is the Levi-Civita symbol, with 
${\hat\epsilon}_{0123}=1$. The quantity $\chi^{ijkl}(x)$ characterizes 
the electromagnetic properties of the vacuum and is as such of 
universal importance.  It is an untwisted tensor density of weight 
$+1$ with 36 independent components. If we decompose it into 
irreducible pieces with respect to the 6-dimensional linear group, 
then we find 
\begin{equation}\label{decomp'}\chi^{ijkl}={}^{(1)}\chi^{ijkl} 
  + {}^{(2)}\chi^{ijkl} + {}^{(3)}\chi^{ijkl}\,, \quad{\rm with} \quad 
  36= 20\oplus 15 \oplus 1\end{equation} independent components, 
respectively. The irreducible pieces of $\chi$ are defined as follows: 
\begin{eqnarray}\label{chiirr2} 
  {}^{(2)}\chi^{ijkl}&:=&\frac{1}{2}\left( \chi^{ijkl}- 
    \chi^{klij}\right)=-{}^{(2)}\chi^{klij}\,,\quad 
  {}^{(3)}\chi^{ijkl}:=\chi^{[ijkl]}\,, \nonumber\\{}^{(1)} 
  \chi^{ijkl}&:=& \chi^{ijkl} -\, 
  {}^{(2)}\chi^{ijkl}-\,{}^{(3)}\chi^{ijkl}={}^{(1)} \chi^{klij}\,. 
\end{eqnarray} 
Since no metric is available, we cannot form traces.  The Abelian
axion piece $^{(3)}\chi^{ijkl}=:\alpha(x)\,\epsilon^{ijkl}$ has been
introduced by Ni \cite{Ni73,Ni77}. Constitutive laws (for matter) with
$^{(2)}\chi^{ijkl}\ne0$ (non-vanishing ``skewon fields''\footnote{The
  name `skewon' has been used earlier in a post-Einsteinian theory of
  gravity by Mann, Moffat, and Taylor \cite{MMT80}, see also Soleng
  and Eeg\cite{SE92}. We are using this name in a different context
  since its old meaning doesn't seem to be in use any longer, see a
  title search in the data banks of SLAC and DESY.} have been
discussed by Nieves and Pal \cite{NP89,NP94}. They yield P- and
CP-violating terms in the field equations. Thus all constitutive
functions in (\ref{decomp'}) can claim respectability from a physical
point of view in the framework of a linear response theory. We will
come back to this question in Sec.3.
 
Our strategy will now be the following: After having specified the 
spacetime relation in (\ref{cl}), we have a predictive theory.  Using 
Hadamard's method \cite{Hadamard,Lichnerowicz} for determining the 
propagation of waves in the spacetime specified so far, namely a 
4-dimensional differentiable manifold carrying the tensor density 
$\chi^{ijkl}$ of (\ref{decomp'}), we will derive, in Sec.2, the {\em 
  Fresnel equation} controlling the waves. It will turn out that the 
axion $\alpha(x)$ drops out and does not affect the waves, in spite of 
showing up in the Maxwell equations. In other words, from 
$\chi^{ijkl}$ in (\ref{decomp'}), only $^{(1)} \chi^{ijkl}$, the 
leading piece, and ${}^{(2)} \chi^{ijkl}$, the skewon field, enter the 
Fresnel equation.  We will give (for the first time) a generally 
covariant and explicit analytical derivation of the corresponding 
Fresnel equation and will show that its wave surfaces are in general 
of {\it quartic} order. 
 
We will study in detail the conditions which must be fulfilled in 
order to factorize the quartic wave surfaces into two equal quadratic 
wave surfaces, the {\it light cone}.  Up to a function, this amounts 
to a new {\it derivation of the metric of spacetime from 
  electromagnetic data}. 
 
In this article, we will restrict ourselves to {\it linear} pre-metric 
electrodynamics. Nevertheless, our methods and results are also useful 
in media with {\it non\/}linear relations between $H$ and $F$, 
provided one studies electromagnetic {\it perturbations} on top of a 
background solution fulfilling the field equations. Then one can 
define, see Appendix A, an effective constitutive tensor density 
$\chi^{ijkl}_{\rm eff}$ that is analogous to $\chi^{ijkl}$ of 
(\ref{decomp'}). In this way, our results, in particular our Fresnel 
equation, also apply to the work done in geometrical optics by De 
Lorenci, Novello, et al.\cite{nov1} in nonlinear electrodynamics 
\`a la Born-Infeld, see also \cite{BLV01,LP00}.

\section{Wave propagation: Fresnel equation}  
                                                                  
We will study the propagation of a discontinuity of the
electromagnetic field following the lines of Ref.\ \cite{OFR00}.  The
surface of discontinuity $S$ is defined locally by a function $\Phi$
such that $\Phi=$~const.\ on $S$. Across $S$, the geometric Hadamard
conditions are satisfied:
\begin{eqnarray} 
&& [F_{ij}] = 0,\qquad [\partial_i F_{jk}] = q_i\, f_{jk},  
\label{had1}\\ 
&& [H_{ij}] = 0,\qquad [\partial_i H_{jk}] = q_i\, h_{jk}.  
\label{had2} 
\end{eqnarray} 
Here $\left[{\cal F}\right](x)$ denotes the discontinuity of a function 
${\cal F}$ across $S$, $q_i:=\partial_i\Phi$ is the wave covector, 
and $f_{ij}$ and $h_{ij}$ are tensors describing the corresponding jumps 
of the derivatives of field strength and excitation. 
 If we use Maxwell's vacuum 
equations $dH =0$ and $ dF =0$, then (\ref{had1}) and (\ref{had2}) 
yield 
\begin{equation}\label{4Dwave} 
  {\epsilon}^{\,ijkl}\, q_{j}\,h_{kl}=0 \,,\qquad 
  {\epsilon}^{\,ijkl}\, q_{j}\,f_{kl}=0\,. 
\end{equation} 
These equations admit non-vanishing wave covectors $q_i$, provided the 
constraints 
\begin{equation}\label{integrability} 
{\epsilon}^{\,ijkl}\, f_{ij}\, f_{kl}=0, \qquad 
{\epsilon}^{\,ijkl}\, h_{ij}\, h_{kl}=0, \qquad 
{\epsilon}^{\,ijkl}\, f_{ij}\, h_{kl}=0 
\end{equation} 
are fulfilled. 
 
We assume that $\chi^{ijkl}$ is continuous across $S$. Then, we find 
{}from (\ref{cl}), (\ref{decomp'}) and (\ref{had1}), (\ref{had2}), 
\begin{equation}\label{clhf} 
h_{ij}=\frac{1}{4}\,{\hat \epsilon}_{ijkl}\,{\chi}{}^{klmn} 
\,f_{mn}\,. 
\end{equation} 
With this spacetime relation for the jumps, the conditions 
(\ref{4Dwave}) can be rewritten as 
\begin{equation}\label{4Dwave2} 
  {\chi}{}^{\,ijkl}\, q_{j}\,f_{kl}= 
  \left(^{(1)}{\chi}{}^{\,ijkl}+\,^{(2)}{\chi}{}^{\,ijkl}\right) 
  q_{j}\,f_{kl}=0\,,\qquad {\epsilon}^{\,ijkl}\, q_{j}\,f_{kl}=0\,, 
\end{equation} 
since $^{(3)}{\chi}{}^{\,ijkl}\, q_{j}\,f_{kl}= 
\alpha\,\epsilon^{ijkl}\, q_{j}\,f_{kl}=0$. Thus the axion field 
$\alpha$ drops out even though it enters the Maxwell equations. 
 
Technically, the system (\ref{4Dwave2}) was analyzed in Ref.\  
\cite{OFR00} in the framework of a 3-(co)vector decomposition of the 
electromagnetic discontinuity tensor $f_{ij}$. Here we present a 
different, generally covariant derivation of the Fresnel equation by 
applying ideas of Tamm \cite{Tamm}. 
 
As a first step, eq.(\ref{4Dwave2})$_2$ is solved by 
\begin{equation} 
f_{ij} = q_ia_j - q_ja_i \,.\label{fqa} 
\end{equation} 
The covector $a_i$ is only defined up to a `gauge' transformation 
\begin{equation}\label{agt} 
a_i\rightarrow a_i + \lambda q_i \,, 
\end{equation}   
with an arbitrary scalar $\lambda$. We substitute (\ref{fqa}) into 
(\ref{4Dwave2})$_1$: 
\begin{equation}\label{4Dwave3} 
  {\chi}{}^{\,ijkl}\,q_{j}q_ka_l=0 \,. 
\end{equation} 
Because of (\ref{agt}), not all the equations in (\ref{4Dwave3}) are 
independent.  In order to isolate the trivial parts of 
(\ref{4Dwave3}), it is convenient to pick a specific (anholonomic) 
{\em coframe} $\vartheta^\alpha=e_i{}^\alpha\, dx^i$, here 
$\alpha,\beta,\ldots=\hat 0, \hat 1, \hat 2, \hat 3$: Namely, we 
identify the zeroth leg of the coframe with the wave covector, that 
is, $\vartheta^{\hat 0}=q$ or, in components, $e_i{}^{\hat 0}=q_i$ or 
$q_\alpha=(1,0,0,0)$. Then 
\begin{equation}\label{eval} 
  {\chi}^{\alpha\beta\gamma\delta}\,q_{\beta}q_\gamma a_\delta= 
  {\chi}^{\alpha\hat{0}\hat{0}\delta}\, a_\delta= 
  {\chi}^{\alpha\hat{0}\hat{0}b}\, a_b=0 \,. 
\end{equation} 
`Spatial' anholonomic indices run over ${a},b,...=\hat 1, \hat 2, \hat 
3$. The zeroth component of (\ref{eval}) vanishes. Thus we find 
\begin{equation}\label{e3} 
  {\chi}^{\hat{0}a\hat{0}b}\, a_b=W^{ab}\,a_b =0\,,\quad{\rm 
    with}\quad W^{ab}:=  {\chi}^{\hat{0}a\hat{0}b}\,. 
\end{equation}

The necessary and sufficient condition for the existence of 
non-trivial solutions for $a_{b}$ is the vanishing of the determinant 
of the $3\times 3$ matrix $W$: 
\begin{equation}\label{det01} 
{\cal W}:=\det W  
=\frac{1}{3!}\epsilon_{ a b c} 
\epsilon_{ d e f}W^{ a d}W^{ b e}W^{ c f}  
=\frac{1}{3!}\epsilon_{ a b c} 
\epsilon_{ d e f} 
{\chi}^{\hat  0  a  \hat 0 d} 
{\chi}^{\hat  0  b  \hat 0 e} 
{\chi}^{\hat  0  c  \hat 0 f} \,. 
\end{equation} 
This yields the Fresnel equation. 
 
We can rewrite $\cal W$ in a fully 4-dimensional covariant manner. 
First, we observe that the 3-dimensional Levi-Civita symbol 
$\epsilon_{ a b c}$ is related to the 4-dimensional one by means of 
$\epsilon_{ a b c}\equiv \epsilon_{\hat 0 a b c}$. Then we can extend 
one $\hat 0$-component of the constitutive tensors to a fourth 
summation index. As a result, we find the identity 
\begin{equation}\label{id01} 
  \epsilon_{\hat 0 a b c} {\chi}{}^{\, \hat 0 a \hat 0 d} {\chi}{}^{\, 
    \hat 0 b \hat 0 e} {\chi}{}^{\, \hat 0 c \hat 0 f}= \epsilon_{\hat 
    0 \beta\gamma\delta} {\chi}{}^{\, \hat 0 \beta \hat 0 d} 
  {\chi}{}^{\, \hat 0 \gamma \hat 0 e} {\chi}{}^{\, \hat 0 \delta \hat 
    0 f}= \frac{1}{2}\, \epsilon_{\alpha\beta\gamma\delta} 
  {\chi}{}^{\, \alpha\beta \hat 0 d} {\chi}{}^{\, \hat 0 \gamma \hat 0 
    e} {\chi}{}^{\, \hat 0 \delta \hat 0 f} \,, 
\end{equation} 
which holds true due to the (anti)symmetry properties of the 
Levi-Civita symbol and of the constitutive tensor. This allows us to 
rewrite (\ref{det01}) as 
\begin{equation}\label{det02} 
{\cal W}=\frac{1}{3!}\,\frac{1}{2}\epsilon_{\alpha\beta\gamma\delta}\, 
\epsilon_{\hat 0  d e f}\, 
{\chi}{}^{\, \alpha\beta \hat 0 d} 
{\chi}{}^{\, \hat 0 \gamma \hat 0 e} 
{\chi}{}^{\, \hat 0 \delta \hat 0 f} \,. 
\end{equation} 
Now we apply the same procedure to the second Levi-Civita symbol and 
finally obtain 
\begin{equation}\label{det03} 
{\cal W}=\frac{1}{4!}\epsilon_{\alpha\beta\gamma\delta}\, 
\epsilon_{\lambda\rho\sigma\tau} 
{\chi}{}^{\, \alpha\beta \hat 0\rho} 
{\chi}{}^{\, \hat 0 \gamma \hat 0\sigma} 
{\chi}{}^{\, \hat 0 \delta \lambda\tau}  \,. 
\end{equation} 
Since $e_i{}^{\hat 0}=q_i$, this can, in coordinate components, be 
written as 
\begin{equation}\label{wgen} 
  {\cal W}=\frac{\theta^2}{4!}\,{\epsilon}_{mnpq}\, 
  {\epsilon}_{rstu}\, {\chi}{}^{\,mnri}\, 
  {\chi}{}^{\,jpsk}\,{\chi}{}^{\,lqtu } \, q_iq_jq_kq_l\, , 
\end{equation} 
with $\theta:=\det(e_i{}^\alpha)$. We define the fourth order tensor 
density of weight $+1$  
\begin{equation}\label{G4}   
  {\cal G}^{ijkl}(\chi):=\frac{1}{4!}\,{\epsilon}_{mnpq}\, 
  {\epsilon}_{rstu}\, {\chi}^{mnr(i}\, {\chi}^{j|ps|k}\, 
  {\chi}^{l)qtu }\,. 
\end{equation} 
It is totally symmetric, ${\cal G}^{ijkl}(\chi)= {\cal 
  G}^{(ijkl)}(\chi)$, and thus has  35 independent components (see 
\cite{Schouten}). With (\ref{G4}), we find the Fresnel equation in the 
generally covariant form 
\begin{equation} \label{Fresnel}   
{\cal G}^{ijkl}(\chi)\,q_i q_j q_k q_l = 0 \,.  
\end{equation} 
 
The general Fresnel equation (\ref{Fresnel}) is always a {\em quartic} 
equation in $q_i$ despite the fact that it was derived from a 
determinant of a $3\times 3$ matrix quadratic in the wave covectors. 
This corrects Denisov and Denisov \cite{DD99} who claim that a 
particular case of the general linear constitutive law may yield a 
sixth order Fresnel equation; in \cite{Lam99}, even a Fresnel equation 
of eighth order is claimed to hold. 
 
The rest of our paper is devoted to finding conditions that will make 
the quartic Fresnel equation {\em factorize} into two quadratic ones 
and to determine under which circumstances the Fresnel equation turns 
out to be a perfect square. 
 
For practical calculations, it is convenient to use a 1+3 coordinate 
decomposition similar to that of Refs.\ \cite{OH99,HOR00,OFR00}. 
Correspondingly, we can rewrite our general result (\ref{Fresnel}) as 
\begin{equation} 
  {\cal W}=q_0^2\left(q_0^4 M + q_0^3q_a\,M^a + q_0^2q_a q_b\,M^{ab} + 
    q_0q_a q_b q_c\,M^{abc} + q_a q_b q_c 
    q_d\,M^{abcd}\right)=0\,,\label{fresnel} 
\end{equation} 
with  
\begin{equation}\label{compare01}  
M:={\cal G}^{0000}\,,\quad M^a:=4{\cal G}^{000a}\,, 
\quad M^{ab}:=6{\cal G}^{00ab}\,,\quad   
\end{equation} 
\begin{equation}\label{compare02} 
M^{abc}:=4{\cal G}^{0abc}\,, \quad  M^{abcd}:={\cal G}^{abcd} \,. 
\end{equation} 
In terms of the $3\times 3$ matrices 
\begin{equation}    
{\cal A}^{ba}:={\chi}{}^{0a0b}\,,   
\qquad {\cal B}_{ba}:=\frac{1}{4}\,{\hat\epsilon}_{acd}\,   
{\chi}{}^{cdef}\,{\hat\epsilon}_{efb}\,,     
\end{equation}    
\begin{equation}    
{\cal C}_{\ a}^{b}:=\frac{1}{2} \,{\hat\epsilon}_{acd}\,    
{\chi}{}^{cd0b}\,,  
\qquad {\cal D}^{\ a}_{b}:=\frac{1}{2} \, {\chi}{}^{0acd}\,  
{\hat\epsilon}_{cdb} \,, 
\end{equation} 
the tensor density $\chi^{ijkl}$ can be written as a $6\times 6$ matrix 
\begin{equation}   
  {\chi}^{IK}= \left( \begin{array}{cl} {\cal B}_{ab}& {\cal D}_a^{\ b} 
\\ 
{\cal C}^a_{\ b} & {\cal A}^{ab} \end{array} \right)\,,\quad{\rm 
with}\quad I,K,...=1,2,...,6\,.\label{almost} 
\end{equation} 
For the $M$'s, we have explicitly: 
\begin{eqnarray}  
M&=&\det{\cal A} \,, \\  
M^a&=& -\hat{\epsilon}_{bcd}\left( {\cal A}^{ba}\,{\cal A}^{ce}\,  
{\cal C}^d_{\ e} + {\cal A}^{ab}\,{\cal A}^{ec}\,{\cal D}_e^{\ d}  
\right)\,,\label{ma1}\\  
M^{ab}&=& \frac{1}{2}\,{\cal A}^{(ab)}\left[({\cal C}^d{}_d)^2 +   
({\cal D}_c{}^c)^2 - ({\cal C}^c{}_d + {\cal D}_d{}^c)({\cal C}^d{}_c +   
{\cal D}_c{}^d)\right]\nonumber\\   
&&+({\cal C}^d{}_c + {\cal D}_c{}^d)({\cal A}^{c(a}{\cal C}^{b)}{}_d  +   
{\cal D}_d{}^{(a}{\cal A}^{b)c}) - {\cal C}^d{}_d  
{\cal A}^{c(a}{\cal C}^{b)}{}_c \nonumber\\   
&& - {\cal D}_c{}^{(a}{\cal A}^{b)c}{\cal D}_d{}^d    
- {\cal A}^{dc}{\cal C}^{(a}{}_c {\cal D}_d{}^{b)}+  
\left({\cal A}^{(ab)}{\cal A}^{dc}-   
{\cal A}^{d(a}{\cal A}^{b)c}\right){\cal B}_{dc}\,,\label{ma2}\\   
M^{abc} &=& \epsilon^{de(c|}\left[{\cal B}_{df}(  
{\cal A}^{ab)}\,{\cal D}_e^{\ f} - {\cal D}_e^{\ a}{\cal A}^{b)f}\,)  
+ {\cal B}_{fd}({\cal A}^{ab)}\,{\cal C}_{\ e}^f -   
{\cal A}^{f|a}{\cal C}_{\ e}^{b)})\,  
\right. \nonumber \\  
&& \left. +{\cal C}^{a}_{\ f}\,{\cal D}_e^{\ b)}\,{\cal D}_d^{\ f}  
+ {\cal D}_f^{\ a}\,{\cal C}^{b)}_{\ e}\,{\cal C}^{f}_{\ d} \right]   
\, ,\label{ma3}\\  
M^{abcd} &=& \epsilon^{ef(c}\epsilon^{|gh|d}\,{\cal B}_{hf}  
\left[\frac{1}{2} \,{\cal A}^{ab)}\,{\cal B}_{ge}  
- {\cal C}^{a}_{\ e}\,{\cal D}_g^{\ b)}\right] \,.\label{ma4}  
\end{eqnarray}  
Here the $M$'s, as totally symmetric tensors in 3 dimensions, carry 
$1\oplus 3\oplus 6\oplus 10 \oplus15=35$ independent components. All 
these results have been verified by using the computer algebra system 
Maple together with the tensor package GrTensor\footnote{See { 
    http://grtensor.org}\,.}. 
 
\section{Axion and skewon} 
 
Before we draw explicitly conclusions from the Fresnel equation 
(\ref{Fresnel}) or (\ref{fresnel}), it is instructive to study the 
structure of ${\cal G}^{ijkl}$. As we saw, this tensor density has 35 
independent components, the same number as 
$^{(1)}\chi^{ijkl}+\,^{(2)}\chi^{ijkl}$. This can be understood as 
follows: Because of (\ref{4Dwave2})$_1$, we have 
\begin{equation}\label{drei} 
{\cal G}^{ijkl}(^{(3)}\chi)=0\,. 
\end{equation}  
This can also be directly verified by substituting $^{(3)}\chi\sim 
\epsilon $ into (\ref{G4}). Accordingly, if the spacetime relation 
(\ref{cl}) contained only the axion piece $^{(3)}\chi$, the 
propagation of electromagnetic waves would not be well-behaved. 
Therefore, in nature, it is excluded that $\chi$ consists merely of 
$^{(3)}\chi\sim \alpha$. On the other hand, if we substitute $\chi$ 
into ${\cal G}$, we find 
\begin{equation} \label{einspluszwei} 
  {\cal G}^{ijkl}(\chi)={\cal G}^{ijkl}(^{(1)}\chi+{}^{(2)}\chi+ 
  {}^{(3)}\chi)=  {\cal G}^{ijkl}(^{(1)}\chi+{}^{(2)}\chi)\,. 
\end{equation}  
Note that $\cal G$ depends on $\chi$ is a cubic way, that is, our last
equation is by no means trivial. Consequently, if either $^{(1)}\chi$
or $^{(2)}\chi$ or both pieces exist in nature --- and this is for
sure --- then the axion piece doesn't interfere with the local laws of
the propagation of electromagnetic waves as determined by the Fresnel
equation. Or expressed in a positive manner: The Abelian axion could
be around, the laws of electrodynamics are compatible with it and the
light propagation wouldn't be affected locally.  There have been
intensive experimental searches for axions, so far without success
\cite{Cooper,Cheng,Stedman}. However, the axion remains a serious
candidate for a particle search in experimental high energy physics.
 
The axion piece has one further interesting property. The kinetic 
electromagnetic energy-momentum current in pre-metric electrodynamics 
reads, see \cite{HO01,HO02}, 
\begin{equation} 
  ^ {\rm k} \Sigma_\alpha :={\frac 1 2}\left[F\wedge(e_\alpha\rfloor 
    H) - H\wedge (e_\alpha\rfloor 
    F)\right]\,.\label{simax}\end{equation} If we substitute the 
spacetime relation (\ref{cl}) into it, we find, in reminiscence of 
(\ref{einspluszwei}), 
\begin{equation}\label{axenergy} 
  ^{\rm k}\Sigma_\alpha(\chi) = \,^{\rm k}\Sigma_\alpha 
  \left(^{(1)}\chi +{}^{(2)}\chi +{}^{(3)}\chi \right)=\,^{\rm 
    k}\Sigma_\alpha \left(^{(1)}\chi +{}^{(2)}\chi \right)\,. 
\end{equation} 
Moreover, $^{\rm k}\Sigma_\alpha(^{(3)}\chi)=0$. Clearly then, the 
axion does not contribute to the electromagnetic energy-momentum 
current. Nevertheless, as Ni \cite{Ni73,Ni77} has shown, it is 
possible to develop a reasonable theory for the Abelian axion. 
Accordingly, the axion piece $^{(3)}\chi$ of the constitutive tensor 
density $\chi$ cannot be dismissed a priori. 
 
Let's turn then to the skewon piece $^{(2)}\chi$. The name we derived 
from the skew- or antisymmetric $6\times 6$ matrix $^{(2)}\chi$ can be 
mapped to. Commonly, this piece is not considered seriously.  The 
conventional argument runs as follows, see Post \cite{Post}.  Suppose 
a Lagrangian 4-form $L$ exists for the electromagnetic field.  In 
general $H\sim \partial L/\partial F$. If $H$ is assumed to be linear 
in $F$, as is done in (\ref{cl}), then $L$ reads $L\sim H\wedge 
F\sim\chi\cdot F\wedge F={}^{(1)}\chi \cdot F\wedge F+ {}^{(3)}\chi \cdot 
F\wedge F$, since the piece with $^{(2)}\chi$ drops out of $L$ because 
of the antisymmetry $^{(2)}\chi^{ijkl}= -\,^{(2)}\chi^{klij}$, see 
(\ref{chiirr2}). Hence, 
\begin{equation}\label{lagrangian} 
  L(\chi)= L(^{(1)}\chi+{}^{(3)}\chi)= L(^{(1)}\chi)+ L(^{(3)}\chi)\,. 
\end{equation} 
The term with $^{(1)}\chi$ eventually becomes the Maxwell Lagrangian, 
the term with $^{(3)}\chi$ part of the axion Lagrangian.  Since we 
conventionally assume that all information of a physical system is 
coded into its Lagrangian, we reject $^{(2)}\chi^{ijkl}\ne 0$ as being 
unphysical. 
 
However, as we saw already in (\ref{axenergy}), if a piece drops out 
from a certain expression, it does not necessarily imply that this 
piece lost its right of existence. Eq.(\ref{lagrangian}) only shows 
that $L$ is `insensitive' to the skewon piece. In contrast, the 
electromagnetic energy-momentum current (\ref{simax}) definitely 
`feels' the contribution from the skewon, 
\begin{equation} \label{skenergy} 
  ^ {\rm k} \Sigma_\alpha(^{(2)}\chi)\ne 0\,, 
\end{equation} 
that is, the skewon piece does carry electromagnetic energy-momentum. 
Thereby, it could displays its presence. Clearly then, we expect that 
$^{(2)}\chi$ influences light propagation. And, indeed, it does. By 
inspection, we find  
\begin{equation}\label{gskenergy} 
  {\cal G}^{ijkl}(^{(1)}\chi+{}^{(2)}\chi)= {\cal G}^{ijkl} 
  (^{(1)}\chi) + \overline{{\cal 
      G}}^{ijkl}(^{(1)}\chi,{}^{(2)}\chi)\,,\end{equation} 
with 
\begin{equation}  \overline{{\cal G}}^{ijkl}(^{(1)}\chi,{}^{(2)} 
  \chi)\ne 0,\label{einsundzw}\end{equation} even though 
\begin{equation} \label{zwei} 
  {\cal G}^{ijkl}(^{(2)}\chi)=0\,. 
\end{equation}  
In other words, like the axion, the skewon cannot propagate light 
decently unless the $^{(1)}\chi$ piece participates. Moreover, 
\begin{equation} \label{zweidrei} 
  {\cal G}^{ijkl}(^{(2)}\chi+{}^{(3)}\chi)=0\,, 
\end{equation}  
i.e., the skewon and the axion piece cannot exist alone or together, 
the `leading' $^{(1)}\chi$ piece is indispensable. If we knew more 
about the detailed structure of $ \overline{{\cal G}}^{ijkl} 
(^{(1)}\chi,{}^{(2)}\chi)$, then we could better judge which modes of 
the skewon are compatible with present-day experiments. 
 
In any case, in linear pre-metric electrodynamics a Lagrangian is not 
assumed to exist a priori. Therefore, it is not alarming that 
$^{(2)}\chi$ drops out from the proto-Lagrangian $L$. Since it carries 
electromagnetic energy-momentum, we will take {\em the possible 
  existence of} $^{(2)}\chi^{ijkl}$ {\em for granted}. 
 
What is then the possible physical meaning of $^{(2)}\chi$? This has 
already been discussed by Nieves and Pal \cite{NP89,NP94}. But let us 
first collect some formalism. Consider a certain vector field 
$\xi=\xi^\alpha e_\alpha$, with the basis $e_\alpha$ of the tangent 
vector space at each point of spacetime. Then we can transvect the 
energy-momentum current (\ref{simax}) with $\xi^\alpha$: 
\begin{equation}\label{QQ} 
  {\cal Q}:=\xi^\alpha\,^{\rm k}\Sigma_\alpha= 
  \frac{1}{2}\left[F\wedge(\xi\rfloor H)- H\wedge(\xi\rfloor 
    F)\right]\,. 
\end{equation} 
The scalar-valued 3-form $\cal Q$ is expected to be related to 
conserved quantities provided we can find suitable (Killing type) 
vector fields $\xi$. Therefore we determine its exterior derivative 
and find after some algebra, 
\begin{equation}\label{dQQ} 
  d{\cal Q}=(\xi\rfloor F)\wedge J+\frac{1}{2}\left( F\wedge {\cal 
      L}_\xi H-H\wedge{\cal L}_\xi F \right)\,, 
\end{equation}  
or, in holonomic components, with ${\cal Q}^i:= 
\epsilon^{ijkl}Q_{jkl}/6$, $\,{\cal J}^i:= \epsilon^{ijkl}J_{jkl}/6$, 
and ${\cal H}^{ij} :=\epsilon^{ijkl}H_{kl}/2$, 
\begin{equation}\label{diqq} 
  \partial_i {\cal Q}^i=\xi^kF_{kl}\,{\cal J}^l+\frac{1}{4}\left( 
    F_{kl}\,{\cal L}_\xi {\cal H}^{kl} -{\cal H}^{kl}\, {\cal L}_\xi 
    F_{kl}\right)\,. 
\end{equation} 
Here ${\cal L}_\xi$ denotes the Lie derivative along the vector $\xi$. 
Now we substitute the linear spacetime relation (\ref{cl}), or ${\cal 
  H}^{kl}=\chi^{klmn}F_{mn}/2$, and find 
\begin{equation}\label{diqqlin} 
  \partial_i {\cal Q}^i=\xi^kF_{kl}\,{\cal J}^l+\frac{1}{8}\left[ 
    F_{kl}\,{\cal L}_\xi (\chi^{klmn}F_{mn}) -\chi^{klmn}F_{mn}\, 
    {\cal L}_\xi F_{kl}\right]\,. 
\end{equation} 
We apply the Leibniz rule of the Lie derivative and rearrange a bit: 
\begin{equation}\label{diqqlin1} 
  \partial_i {\cal Q}^i=\xi^kF_{kl}\,{\cal J}^l+\frac{1}{8}\left[ 
    ({\cal L}_\xi \chi^{ijkl})\,F_{ij}F_{kl}+(\chi^{ijkl}- 
    \chi^{klij})\,F_{ij}\, {\cal L}_\xi F_{kl}\right]\,. 
\end{equation} 
We substitute the irreducible pieces of $\chi^{ijkl}$. Then we have 
\begin{equation}\label{diqqlin2} 
  \partial_i {\cal Q}^i=\xi^kF_{kl}\,{\cal J}^l+\frac{1}{8}\, {\cal 
    L}_\xi \left(^{(1)}\chi^{ijkl}+\,^{(3)}\chi^{ijkl} 
  \right)\,F_{ij}F_{kl}+ \frac{1}{4}\,^{(2)}\chi^{ijkl} \,F_{ij}\, 
  {\cal L}_\xi F_{kl}\,. 
\end{equation} 
 
Here it is manifest that $^{(1)}\chi$ and the axion $^{(3)}\chi$ 
behave qualitatively different as compared to the skewon $^{(2)}\chi$. 
If $^{(1)}\chi$ and $^{(3)}\chi$ carry a symmetry along $\xi$ such 
that $ {\cal L}_\xi{}^{(1)}\chi^{ijkl}=0$ and $ {\cal L}_\xi 
{}^{(3)}\chi^{ijkl}=0$, then, in vacuum, i.e., for ${\cal J}^i=0$, we 
still have {\em non-}conservation because of the offending term 
$^{(2)}\chi\, F\dot{ F}$. Here the dot symbolizes the Lie derivative 
along $\xi$. If $\xi$ can be interpreted as a `time' direction, then 
$\cal Q$ represents the electromagnetic energy density. 
 
In any case we see that $^{(2)}\chi$ induces a {\em dissipative} term 
with a first `time' derivative. This is what we might have expected 
since, in general, dissipative phenomena cannot be described 
conveniently in a Lagrangian framework.  It is then our hypothesis 
that the skewon piece $^{(2)}\chi$ can represent a field which is {\em 
  odd under T transformations}. This is also what had been discussed 
by Nieves and Pal \cite{NP89,NP94}. Of course, we must investigate how 
this {\sl skewon}, as we may call it in a preliminary way, disturbs 
the light cone and whether there is perhaps only a viable subclass of 
the 15 independent components of the skewon. An isotropic skewon $s$ 
could be constructed by putting in (\ref{almost}) ${\cal 
  C}^a{}_b=s\,\delta^a_b$ and ${\cal D}_a{}^b=-s\,\delta_a^b$, see 
Nieves and Pal \cite{NP94}. 
 
Let us try then to collect our results in order to get a rough picture 
of the physical meaning of these different irreducible pieces. The 
piece $^{(1)}\chi^{ijkl}$ is indispensable for an appropriate 
propagation of light, as we saw in (\ref{drei}), (\ref{zwei}), and 
(\ref{zweidrei}).  Thus we call it the leading piece of $\chi^{ijkl}$. 
 
If $^{(1)}\chi^{ijkl}$ exists alone and for the spacetime relation 
(\ref{cl}) {electric/magnetic reciprocity} is assumed to hold 
additionally, see Sec.4, then, up to an unknown function, the metric 
of spacetime can be derived including its Lorentzian signature 
\cite{OH99,HOR00,GR01}.  One can think of this reduction in the way 
that electric/magnetic reciprocity cuts the 20 components of 
$^{(1)}\chi^{ijkl}$ into half, that is, only 10 components are left 
for the metric.  Modulo an undetermined function, we have then 9 
remaining components. These 9 remaining components of 
$^{(1)}\chi^{ijkl}$ determine the {\em light cone} at each point of 
spacetime. Accordingly, in $^{(1)}\chi^{ijkl}$ the light cone of 
spacetime is hidden and thereby conventional Maxwell-Lorentzian vacuum 
electrodynamics as well. To put it more geometrically, the first 
irreducible piece $^{(1)}\chi$, via electric/magnetic reciprocity, 
yields the {\em conformal} structure of spacetime. In this sense, 
there is no doubt that $^{(1)}\chi^{ijkl}$ is the principal part of 
the constitutive tensor density $\chi$ of the vacuum. But, as we 
argued above, there are no a priori reasons for excluding either the 
axion $^{(3)}\chi$ or the skewon $^{(2)}\chi$. In particular, the 
Fresnel equation and the electromagnetic energy-momentum current can 
accommodate both pieces in a plausible way.

 
\section{Almost complex structure}\label{complex}  
 
Let us now search for a condition restricting the constitutive tensor 
density $\chi^{ijkl}$ of the spacetime relation (\ref{cl}). Since the 
times of Maxwell and Heaviside, in the equations of electrodynamics a 
certain symmetry was noticed between the electric and the magnetic 
quantities and was used in theoretical discussions. We formulate 
electric/magnetic reciprocity as follows \cite{HO02}: Given the 
energy-momentum current (\ref{simax}). It is electric-magnetic 
reciprocal, i.e., it remains invariant $\Sigmakin\rightarrow 
\Sigmakin$ under the {transformation} 
\begin{equation}\label{duality1} 
  H\rightarrow \zeta F\,,\quad F\rightarrow 
  -\frac{1}{\zeta}\,H\,, 
\end{equation} 
with the dimensionful pseudo-scalar function $\zeta=\zeta(x)$.  
 
We now postulate {\em electric/magnetic reciprocity of the spacetime 
relation (\ref{cl})}. Under (\ref{duality1}), the spacetime relation 
transforms into 
\begin{equation}\label{duality2}   
  \zeta\,F_{ij}=-\frac{1}{4\,\zeta}\,{\hat 
    \epsilon}_{ijkl}\,\chi^{klmn}\,H_{mn} \,. 
\end{equation}   
If we substitute this into (\ref{cl}), we find, after some 
algebra, the consistency condition 
\begin{equation}\label{duality3} 
  -\frac{1}{8\,\zeta^2}\,({\hat \epsilon}_{ijmn}\,\chi^{mnpq})\,( {\hat 
    \epsilon}_{pqr\!s}\,\chi^{r\!skl})=\delta^{kl}_{ij}\,, 
\end{equation} 
with the function 
\begin{equation}\label{duality4} 
  \zeta^2:= -\frac{1}{96}\,({\hat \epsilon}_{ijmn}\,\chi^{mnpq})\,( 
  {\hat \epsilon}_{pqr\!s}\,\chi^{r\!sij})\,. 
\end{equation} 
In order to allow for a more compact notation, we introduce the 
dimensionless density $\stackrel{\rm 
  o}{\chi}{\!}^{ijkl}:={\chi}^{ijkl}/\zeta$ and define  
\begin{equation}\label{circle} 
  \kappa_{ij}{}^{kl}=\frac{1}{2}\,{\hat \epsilon}_{ijmn}\,\stackrel{\rm 
    o}{\chi}{\!}^{mnkl}\,. 
\end{equation} 
Then the {\em closure relation} reads 
\begin{equation}\label{duality5} 
  \kappa_{ij}{}^{mn} \kappa_{mn}{}^{kl}=-2\delta_{ij}^{kl} 
\end{equation} 
or, even more compactly, as $6\times 6$ matrix equation, 
\begin{equation}\label{close1a} 
{\kappabf}\,{\kappabf}=-\,\mathbf{1}_6. 
\end{equation} Mathematically this means that the operator  
${\kappabf}$ represents an almost complex structure on the space of 
2-forms.  For the special case of $^{(2)}\chi={}^{(3)}\chi\equiv 0$, 
such closure relations were first discussed by Toupin \cite{Toupin}, 
Sch\"onberg \cite{Schoenberg}, and Jadczyk \cite{Jadczyk}. 
 
Let us now make the closure relation explicit. We turn back to the 
constitutive $6\times 6$ matrix (\ref{almost}). We define 
dimensionless $3\times 3$ matrices $\stackrel{\;{\rm o}}{\cal 
  A}\;:={\cal A}/\zeta$, etc. In terms of these dimensionless matrices 
(we immediately drop the small circle for convenience), the closure 
relation reads, 
\begin{eqnarray}  
{\cal A}^{ac}{\cal B}_{cb} +  {\cal C}^a{}_c{\cal C}^c{}_b &=&   
- \delta^a_b,\label{almostclose1} \\  
{\cal C}^a{}_c {\cal A}^{cb} + {\cal A}^{ac}{\cal D}_c^{\ b}   
&=& 0,\label{almostclose2}\\  
{\cal B}_{ac}{\cal C}^c{}_b + {\cal D}_a^{\ c}{\cal B}_{cb}    
&=& 0, \label{almostclose3}\\  
{\cal B}_{ac}{\cal A}^{cb} +  {\cal D}_a^{\ c}{\cal D}_c^{\ b}   
&=& -\delta^b_a.\label{almostclose4}  
\end{eqnarray}  
Assume $\det {\cal B}\neq 0$. Then we can find the general 
non-degenerate solution. We define the matrix $K_{ab}$ by 
\begin{equation}  
  K:={\cal BC}\,,\qquad{\rm i.e.}\qquad {\cal C}= {\cal 
    B}^{-1}K\,,\label{CDKK} 
\end{equation}  
and substitute it into (\ref{almostclose3}), 
\begin{equation}  
{\cal D} = - K{\cal B}^{-1}.\label{KK}  
\end{equation}  
Next, we solve (\ref{almostclose1}) with respect to $\cal A$: 
\begin{equation}  
{\cal A}=-{\cal B}^{-1}-{\cal B}^{-1}K{\cal B}^{-1}K{\cal  
B}^{-1}.\label{ABK}  
\end{equation}  
We multiply (\ref{ABK}) by $\cal C$ from the left and by $\cal D$ from 
the right, respectively, and find with (\ref{CDKK}) and (\ref{KK}), 
\begin{eqnarray}  
{\cal CA} &=& - {\cal B}^{-1}K{\cal B}^{-1}   
- {\cal B}^{-1}K{\cal B}^{-1}K{\cal B}^{-1}K{\cal B}^{-1},\\  
{\cal AD} &=& + {\cal B}^{-1}K{\cal B}^{-1}   
+ {\cal B}^{-1}K{\cal B}^{-1}K{\cal B}^{-1}K{\cal B}^{-1}.  
\end{eqnarray}  
Thus, (\ref{almostclose2}) is automatically satisfied.  
Accordingly, only (\ref{almostclose4}) has still to be checked. We compute its
first and second term of its left side, 
\begin{eqnarray}  
{\cal BA} &=& -1 - K{\cal B}^{-1}K{\cal B}^{-1}\,,\\  
{\cal D}^2 &=& K {\cal B}^{-1} K {\cal B}^{-1}\,, 
\end{eqnarray} and find that it is fulfilled, indeed. 
  
Summing up, we have derived the general solution of the closure 
relation (\ref{close1a}) in terms of two arbitrary matrices ${\cal B}$ 
and $K$ as 
\begin{eqnarray}  
{\cal A} &=& - {\cal B}^{-1} - {\cal B}^{-1}K{\cal B}^{-1}K{\cal  
B}^{-1}\,,\\  
{\cal C} &=& {\cal B}^{-1}K\,,\\  
{\cal D} &=& - K{\cal B}^{-1}\,.  
\end{eqnarray}  
The solution thus has $2\times 9 = 18$ independent components. 
Alternatively, one can write the solution as 
\begin{eqnarray}  
{\cal A} &=&
- (1+{\cal C}^2){\cal B}^{-1}\, ,\\ {\cal D} &=& - {\cal BCB}^{-1}\,, 
\end{eqnarray}  
which is parametrized by the arbitrary matrices ${\cal B}$ and ${\cal C}$  
with altogether 18 independent components.  
  
\section{Spacetime metric}\label{metric}  
  
Can we construct the spacetime metric by using the new general solution   
for the generalized closure relation? Technically, this is equivalent  
to the question: Can the general Fresnel equation be reduced to the   
light-cone structure? Here we study the conditions for such a reduction.   
  
Let us decompose the arbitrary matrix ${\cal B}$ into its symmetric and  
antisymmetric parts,  
\begin{equation}  
  {\cal B}_{ab} = b_{ab} + \hat{\epsilon}_{abc}n^c\,,\quad{\rm 
    with}\quad b_{ab}:= {\cal B}_{(ab)}\,,\quad n^c:= \epsilon^{cab} 
  \,B_{[ab]}\,.\label{Bn} 
\end{equation}  
Note that $ b_{ab}$ contributes to $^{(1)}\chi$ and $n^c$ to 
$^{(2)}\chi$.  Now we can lower the index of $n^a$ by means of 
$b_{ab}$, namely, $n_a:=b_{bc}\,n^c$ and $n^2 := n^cn_c = 
b_{ab}\,n^an^b$, and, provided $\det b\ne 0$, we can raise an index by 
$b^{ab}$ which denotes the inverse of $b_{ab}$. We find $\det {\cal B} 
= \det b + n^2$, and the inverse of (\ref{Bn}) reads 
\begin{equation}\label{invBn}  
{\cal B}^{ab} = {\frac 1 {\det{b}+n^2}}\left({\bar b}^{ab} + n^an^b -   
\epsilon^{abc}\,n_c\right).  
\end{equation}  
Here the symmetric matrix ${\bar b}^{ab}$ is the matrix of the minors 
of $b_{ab}$. If $\det b\ne 0$, then ${\bar b}^{ab}=b^{ab}\det{b}$. 
  
Let us find out what is qualitatively new in the asymmetric case as 
compared to the previously studied symmetric case \cite{OFR00}. For 
this purpose we consider the particular solution of the closure 
relation for $K=0$. Consequently, ${\cal C} = {\cal D} =0$ and ${\cal 
  A} = - {\cal B}^{-1}$. Then, by substituting (\ref{Bn}) into 
(\ref{fresnel}), we obtain: 
\begin{equation}  
{\cal W} = -\,{\frac {q_0^2}{(\det{b}+n^2)}}\left[q_0^4  
- 2q_0^2\left(q^2\,\det{b} - (qn)^2\right) + \left(q^2\,\det{b} +   
(qn)^2\right)^2 \right]. \label{freC0}  
\end{equation}  
Here we used the abbreviation $(qn):=q_a\,n^a$.   
  
In general, this expression is neither a square of a quadratic 
polynomial nor a product of two quadratic polynomials. In other words, 
neither a light cone nor a birefringence (double light cone) structure 
arises generically. In order to study the reduction conditions, let us 
assume that the Fresnel equation is a product of two quadratic 
equations for $q_i$, i.e., the spacetime `medium' is birefringent. 
Accordingly, for (\ref{freC0}) we make the general ansatz 
\begin{equation}  
{\cal W} = -\,\frac{q_0^2}{(\det{b}+n^2)} (q_0^2-\alpha)(q_0^2-\beta)=  
-\frac{q_0^2}{(\det{b}+n^2)}\left[q_0^4 - (\alpha+\beta) q_0^2  
+\alpha\beta   
\right], \label{freC02}  
\end{equation}  
with some polynomials $\alpha$ and $\beta$ of order 2 in $q_a$.  This  
implies the relations  
\begin{equation}\label{eqab}  
\alpha+\beta=2\left({\bar q}^2 - (qn)^2\right) , \qquad   
\alpha\beta= \left({\bar q}^2 + (qn)^2\right)^2 \,,  
\end{equation}  
with $\bar{q}^2 := q_a q_b \bar{b}^{ab}$. Since $\alpha$ and $\beta$  
enter symmetrically in (\ref{eqab}), the solutions of this nonlinear  
system can be given in the form  
\begin{equation}\label{solalphabeta}  
\alpha=  -\,\left[(qn) +\sqrt{-\overline{q}^2}\right]^2 , \qquad   
\beta= -\,\left[(qn) - \sqrt{-\overline{q}^2}\right]^2 \, .  
\end{equation}  
  
Thus, the question of the reducibility of the Fresnel equation  
translates into the algebraic problem of whether the square root  
$\sqrt{-\overline{q}^2}$ is a real linear polynomial in $q_a$. There  
are three cases, depending on the rank of the $3\times 3$ matrix  
$b_{ab}$.  
  
(i) When $b_{ab}$ has rank 3, in other words, when $\det b \neq 0$, 
then we can write $\overline{q}^2 = q_aq_b \,b^{ab}\det b$, and the 
general conclusion is that {\it no factorization into light-cones} is 
possible (the roots $\alpha$ are complex), unless $n^a=0$. This latter 
condition implies that the constitutive tensor is symmetric, and the 
results of \cite{OH99,OFR00,GR01} are recovered. 
  
(ii) When $b_{ab}$ has rank 2, i.e., $\det b =0$, but at least one of the   
minors is nontrivial. Then, without loss of generality, we can assume the   
following structure of the matrix $b$:  
\begin{equation}  
b_{ab} = \left(\begin{array}{ccc}b_{11}&b_{12}&0\\ b_{12}&b_{22}& 0\\  
0 & 0 & 0\end{array}\right).\label{b-deg}  
\end{equation}  
Its only non-vanishing minor is   
\begin{equation}  
\overline{b}^{33} = \left|\begin{array}{cc}b_{11}&b_{12}\\ b_{21}&b_{22}  
\end{array}\right| = b_{11}b_{22}-b_{12}^2 \neq 0.  
\end{equation}  
Note that (\ref{b-deg}) is the most general form of a rank 2 matrix 
$b_{ab}$, up to a renaming of the coordinates. In order to avoid 
complex solutions, we have to assume that the minor $\overline{b}^{33} 
= -\,\mu^2 < 0$, so that $\sqrt{-{\bar q}^2}=\mu q_3$. Then 
(\ref{solalphabeta}) leads to 
\begin{equation}  
\alpha = -\,\left[q_1n^1 + q_2n^2 + q_3(n^3+\mu)\right]^2, \qquad   
\beta = -\,\left[q_1n^1 + q_2n^2 + q_3(n^3-\mu)\right]^2\,.  
\end{equation}  
The interpretation is clear: we have birefringence, i.e., two  
light-cones. In this case, the Fresnel equation is found to be  
\begin{eqnarray}  
{\cal W} = -\,{\frac {q_0^2}{b_{11}(n^1)^2+2b_{12}n^1n^2+b_{22}(n^2)^2}}  
\,&& (q_0^2+\left[q_1n^1 + q_2n^2 + q_3(n^3+\mu)\right]^2) \nonumber \\  
&&\times (q_0^2+\left[q_1n^1 + q_2n^2 + q_3(n^3-\mu)\right]^2)=0\,.  
\end{eqnarray}  
Then we can read off, up to conformal factors, the components of the  
two corresponding `metric' tensors defining the light-cones:  
\begin{equation}  
g_1^{ij}=\left(   \begin{array}{cccc}   
                1 & 0 & 0 & 0 \\   
                0 & (n^1)^2 & n^1 n^2 & n^1(n^3+\mu) \\  
                0 & n^1n^2 & (n^2)^2 & n^2n^3 \\  
                0 & n^1(n^3+\mu) & n^2n^3 & (n^3+\mu)^2   
         \end{array}\right) \, ,  
\end{equation}  
\begin{equation}  
g_2^{ij}=\left(   \begin{array}{cccc}   
                1 & 0 & 0 & 0 \\   
                0 & (n^1)^2 & n^1 n^2 & n^1(n^3-\mu) \\  
                0 & n^1n^2 & (n^2)^2 & n^2n^3 \\  
                0 & n^1(n^3-\mu) & n^2n^3 & (n^3-\mu)^2   
         \end{array}\right) \, .  
\end{equation}  
We can verify that $\det (g_1^{ij})=\det (g_2^{ij})=  
(n^1)^2(n^2)^2{\bar{b}}^{33}=-(n^1)^2(n^2)^2\mu^2<0$, so that both metrics   
have the correct Lorentzian signature.  
  
(iii) When the $3\times 3$ matrix $b_{ab}$ has rank 1: In this case 
all the minors are zero, i.e., $\overline{q}^{ab}=0$, which 
corresponds to the case 2 for $\mu=0$. We then see that the Fresnel 
equation reduces to a single light cone, but the resulting metric is 
degenerated, since $\det (g^{ij})=0$. 
  
\section{Discussion and conclusion}  
  
In this paper we extended appreciably our earlier results 
\cite{OH99,HOR00,OFR00,GR01} of linear pre-metric electrodynamics by 
relaxing the symmetry and the closure property of the constitutive 
tensor $\chi^{ijkl}$ of spacetime. In the most general case of a 
linear spacetime relation (when both {\em closure} and {\em symmetry} 
are absent), the wave propagation is governed by the newly derived 
Fresnel type equation (\ref{Fresnel}) which is still of quartic order, 
contrary to the claims in Ref.\ \cite{DD99}. We have deduced the 
tensor density ${\cal G}^{ijkl}$ in (\ref{G4}) which induces the 
generally covariant form of the Fresnel equation (\ref{Fresnel}). We 
studied some properties of ${\cal G}^{ijkl}$ and showed that 
$^{(1)}\chi^{ijkl}$, as the leading part of $\chi^{ijkl}$, is  
indispensable for a decent propagation of electromagnetic waves. 
However, also each of the two remaining parts, the axion 
$^{(3)}\chi^{ijkl}$ and the skewon $^{(2)}\chi^{ijkl}$, has a 
well-defined physical meaning. The axion drops out from light 
propagation. High energy physicist search for such type of particles. 
The skewon seems to be T-odd and related to dissipative processes. Its 
influence on light propagation, see (\ref{einsundzw}), deserves 
further study. 
  
In Sec.\ref{complex}, the general solution of the closure relation is 
presented for the case of an asymmetric linear $\chi^{ijkl}$. Within a 
particular class of the asymmetric solutions of the closure relation, 
the reduction of the Fresnel equation was analyzed in detail. We have 
demonstrated that some of these solutions can yield birefringence. A 
preliminary study of the general case with $K\neq 0$ shows that the 
conclusions remain qualitatively the same as those presented in 
Sec.\ref{metric}. 
  
Thus, we can conclude that the conditions of closure and symmetry of 
$\chi^{ijkl}$ are sufficient for the existence of a well-defined light 
cone structure. If any of these conditions is violated, the light cone 
structure seems to be lost. The necessary conditions have still to be 
found. Of course, if one wants to study possible violations of Lorentz 
invariance by means of the skewon $^{(2)}\chi^{ijkl}$, e.g., then the 
light cone structure cannot be considered as sacrosanct any longer. 
\medskip 

\noindent {\bf Acknowledgments}  
  
GFR would like to thank the German Academic Exchange Service (DAAD)
for a graduate fellowship (Kennziffer A/98/00829). YNO is grateful to
the Alexander von Humboldt Foundation for support.

\appendix 
 
\section{An effective constitutive tensor}\label{ab} 
 
Consider an arbitrary local spacetime relation $H=H(F)$. In 
components, the corresponding Maxwell equation, for $J=0$, reads 
\begin{equation}\label{mehc} 
\epsilon^{ijkl}\,\partial_j\, H_{kl}=0 \,. 
\end{equation} 
A small perturbation $\Delta F$ of the electromagnetic field around 
some background $\bar F$ can be written as $F={\bar F}+\Delta F$. 
Then, to first order in the perturbation, we have for the excitation 
\begin{equation}\label{expH} 
  H_{kl}(F)=H_{kl}({\bar F})+\frac{1}{2} \left.\frac{\partial 
      H_{kl}}{\partial F_{mn}}\right|_{\bar F} \Delta F_{mn} \,. 
\end{equation} 
Inserting (\ref{expH}) into (\ref{mehc}) and assuming that the 
background field $\bar F$ is a solution of (\ref{mehc}), i.e. 
$\epsilon^{ijkl}\,\partial_j\,H_{kl}({\bar F})=0$, we obtain an 
equation for the perturbation: 
\begin{equation} 
  \partial_j\left(\chi^{ijkl}_{\rm eff} \Delta F_{kl}\right)=0 \,,\qquad 
  \chi^{ijkl}_{\rm eff}:=\frac{1}{2}\epsilon^{ijmn} 
  \left.\frac{\partial H_{mn}}{\partial F_{kl}}\right|_{\bar F} \,. 
\end{equation} 
The effective constitutive tensor density $\chi^{ijkl}_{\rm eff}$ 
will, in general, depend on the local constitutive law and on the 
background field $\bar F$.

\end{document}